\documentclass[twocolumn,showpacs,preprintnumbers,amsmath,amssymb]{revtex4}
\usepackage{graphicx}% Include figure files
\usepackage{dcolumn}% Align table columns on decimal point
\usepackage{bm}% bold math

\newcommand{\frat}[2]{\frac{\textstyle #1}{\textstyle #2}}
\newcommand{\vf}[1]{\mbox{\boldmath $#1$}}
\newcommand{\nomer}[1]{\mbox{$\cal N$\hspace{-.5ex}\raisebox{.3ex}
           {\underline{\tiny 0}$\!$} #1}}

\begin{document}

\title{Quark ensembles with infinite correlation length
}

\author{S. V. Molodtsov}
 \altaffiliation[Also at ]{
Institute of Theoretical and Experimental Physics, Moscow, RUSSIA}
\affiliation{%
Joint Institute for Nuclear Research, Dubna,
Moscow region, RUSSIA%\textbackslash\textbackslash
}%
\author{G. M. Zinovjev}
% \homepage{http://www.Second.institution.edu/~Charlie.Author}
\affiliation{
Bogolyubov Institute for Theoretical Physics,
National Academy of Sciences of Ukraine, Kiev, UKRAINE
}%

\date{\today}% It is always \today, today,
             %  but any date may be explicitly specified

\begin{abstract}
By studying quark ensembles with infinite correlation length we formulate
 the quantum field theory model that, as we show, is exactly integrable and develops
an instability of its standard vacuum ensemble (the Dirac sea). We argue such an
instability is rooted in high ground state degeneracy (for 'realistic' space-time dimensions)
featuring a fairly specific form of energy distribution, and with the cutoff parameter
going to infinity this inherent energy distribution becomes infinitely narrow and leads
to large (unlimited) fluctuations. Analysing some possible vacuum ensembles such as
the Dirac sea, neutral ensemble, color superconducting and BCS states we find out
the strong arguments in favor of the BCS state as the ground state of color interacting
quark ensemble.
\end{abstract}

     % PACS, the Physics and Astronomy
                              % Classification Scheme.
%\keywords{Suggested keywords}%Use showkeys class option if keyword
                              %display desired
\maketitle

\section{Introduction}
%1 2 3 4 5
Investigating quark ensembles with an infinite correlation length looks, from the view point of
 fundamental strong interactions existing between ensemble elements,
 like a purely academic task because the successful practical field theory (QCD) develops finite
 inherent scale as an abundant experience of phenomenological studies (and lattice QCD theory) gives
 an evidence. However, a true nature of these scale
is still pretty uncertain problem \cite{fad}. Here we are concerned with
 this problem by searching for the indicative consequences if the proper scale has already raised in
 a quantum field model. In our particular case of strongly interacting fields a model with 'infinite'\
 correlation length might  be understood as one in which a size is determined by
 the characteristic vacuum box ($L\sim \Lambda^{-1}_{\mbox{\scriptsize{QCD}}}$).
  Moreover, we show, in what
 follows, the consideration of respective quark ensembles is getting essential technical simplification
 because the field theory models of our interest prove to be exactly integrable
(in the sense by Thirring or Luttinger). This remarkable property of certain
class of field theoretical models makes possible to proceed substantially
beyond the perturbative approximation and plays role of great importance in
 understanding the principal problems of quantum field theory \cite{tir}, \cite{bik}. Besides, these
 models (called further as the KKB models) are also well-known and fruitful in the context of
 condensed matter physics \cite{kkb}.
We believe that studying
 the fundamental ensemble features is quite relevant and insightful to deal.

% 6, 7, 8, 9, 10, 11, 12, 13
In fact, this feature of exact integrability has already been exploited \cite{mz} at comparative
analysis of the KKB and Nambu-Jona-Lasinio (NJL) models (both are the models of four-quark
interaction). The model with 'infinite'\ correlation length (KKB) provides also
an interesting possibility to evaluate a role of quantum correlations only,
i.e. at absence of a customary impact force intrinsic in classical dynamics
or electrodynamics,
%%%%%%%%%%%%%%%%%%%%%
 what is quite inherent in studying a system of fermions and regarded
as an exchange force \cite{atom},
\cite{ato}.
%%%%%%%%%%%%%%%%%%%%%
Apparently, in this connection the problem of treating
a system response to any external influence appears to be of
special interest, for example, at analysing a system behavior in an external fields.

Ensemble action in which we are interested to investigate is presented as
\begin{equation}
%1
\label{1}
\hspace{-0.25cm}S=\!\!\!\int dt d^{\mbox{\scriptsize{D}}}
x\left[\bar q(i \hat\partial-m)q-\frat{g}{2}~j^a_\mu
\int d^{\mbox{\scriptsize{D}}}y~j'^a_\mu~F(x,y)\right].
\end{equation}
Here $j^a_\mu=\bar qt^a\gamma_\mu q$ is a quark current with respective quark
field operators $q$, $\bar q=q^+\gamma_0$ taken at the spatial point $x$
(the primed variables refer to the point $y$), $m$ is the current quark mass,
$t^a=\lambda^a/2$ are the generators
of the $SU(N_c)$ color gauge group, $\mu=0,1,2,3,\dots D$ and $\hat\partial$
are the partial derivatives over time $t$ and coordinates $x$ spanned on
the corresponding $\gamma$-matrices.
The form of color interaction plays a significant role below, but we start
our discussion with considering simpler abelian version. In two-dimensional
(time and one spatial component, $D=1$) formulation such an ensemble
corresponds
to the Thirring or Luttinger model \cite{tir}.

For the sake of simplicity the form-factor $F(x,y)$ is put to be translation
invariant, $F(x,y)=F(x-y)$, and dimensionless with singling out a proper
constant $g$.
There is no any preferable spatial point in the ensemble with 'infinite'\
correlation length
(a force which is usually defined as a gradient of potential equals to zero),
and it is just
what we have in mind, while talking about absence of the customary force
interaction. In
principle, the form-factor could be taken as $F(x)=1$ and then the
corresponding
Fourier-transformation $\widetilde F(p)=\int
d^{\mbox{\scriptsize{D}}}x~e^{ipx}F(x)=\delta(p)$
has a dimension $[F(p)]=L^{\mbox{\scriptsize{D}}}$. Concurrently, the
dimension of quark
fields comes about $[q]=L^{\mbox{\scriptsize{-D/2}}}$ and the coupling
constant becomes
dimensional $[g]=L^{-1}$. In what follows, we are dealing with the densities
of 'measurable'\ quantities, for example, an energy density
${\cal E}=E/L^{\mbox{\scriptsize{D}}}$ where $E$ is a total
ensemble energy.
In order to simplify the formulae we do not include the factors
$L^{\mbox{\scriptsize{-D/2}}}$, in the definition of fermion fields, because
they can
be easily recast if necessary and as to the observables they are present via
the corresponding factor of box volume $V=L^{\mbox{\scriptsize{D}}}$. What is
more
specific feature of interaction form considered concerns a formal absence of
scattering,
i.e. quark incoming momentum coincides with the quark outgoing momentum in
the scattering
process.

It should, perhaps, be recalled how, in principle, the effective
form of interaction Eq. (\ref{1}) could be obtained from QCD.
It is assumed that quarks are under the influence of
 strong stochastic vacuum gluon fields. Then using the coarse-graining
procedure
for quark ensemble in quasistationary state we obtain product of interesting us quark currents
associated with corresponding correlator (condensate) of the gluon field
$\langle A^{a}_\mu A'^{b}_\nu\rangle$.
In simplest form it is a color singlet.
For sake of simplicity we restrict ourselves by a contact interaction in time
(without retardation) and, hence, we do not include corresponding delta-function in time in
the form of form-factor
\begin{equation}
%2
\label{cor}
\langle A^{a}_\mu A'^{b}_\nu\rangle=G~\delta^{ab}~\delta_{\mu\nu}~F({\vf x}-
{\vf y})~.
\end{equation}
Of course, there are another terms, spanned on the vector of relative distance
$x-y$.
It is clear that this simplest correlation function is only one of the  fragments
of corresponding ordered exponent and
the four-fermion interaction is clearly accompanied by infinite number of
multifermion vertices \cite{sim}.

\section{Two-dimensional model}
As a matter of fact, the two-dimensional version of model (\ref{1}) is very
well known and has been intensively studied for almost fifty years creating a
fundamental
basis for several areas in many-fermion physics research with the exciting
claims of exact and
complete integrability \cite{tir}. Even nowadays this model is practically
the only
reliable instrument to describe (beyond the one-particle perturbation theory)
a system response
to an external influence.
% 15
It seems curious, the case of correlations dominance, that we are interested
in,
was systematically excluded from consideration by applying the subtraction
procedure,
rejecting the contribution $\widetilde F(0)$.
This variant was not thoroughly investigated because of
various reasons, one of which is certainly too much simple.

If we limit ourselves to the abelian form of interaction (dropping the group
generators $t^a$ out) it is possible to introduce a corresponding doublet of
fields $q=(q_1,q_2)$
(instead of a full-fledged particle spin) for the two-dimensional model
of Eq. (\ref{1}). Then
taking the $\gamma$-matrices in the form  $\gamma_0=\sigma_2$,
$\gamma_1=i\sigma_1$,
where $\sigma_1$, $\sigma_2$ are the Pauli  matrices we
are able to transform the Lagrangian density to the form typical for models
of Ref. \cite{tir}
\begin{equation}
%3
\label{3}
{\cal L}=\bar q(i \hat\partial-m)q-g~(q^+_1 q_1~q'^+_2 q'_2+
q^+_2 q_2~q'^+_1 q'_1)~.
\end{equation}
In order not to overload formulae we omit the integration over coordinate
$y$, but keep it in mind putting the primes over the corresponding fermion
fields in the
proper places. The Hamiltonian density of system under consideration
can be written down in the form
\begin{eqnarray}
%4
\label{4}
{\cal H}&=&q^+_1~p~q_1-q^+_2~p~q_2+m(q^+_1 q_2+q^+_2 q_1)+\nonumber\\
[-.21cm]\\[-.2cm]
        &+&g~(q^+_1 q_1~q'^+_2 q'_2+q^+_2 q_2~q'^+_1 q'_1)~,\nonumber
\end{eqnarray}
where $p = i\partial_x$. Now following Thirring  \cite{tir} we introduce a reference
vacuum state $|0)$ featuring the components of fermion fields as
$q_{1x}|0)=q_{2x}|0)=0$.
Since we consider a system into a finite size box, this condition is
assumed to be valid in the
corresponding discrete spatial points for the fermions with antiperiodic
boundary
conditions. First, we consider the system in the chiral limit ($m=0$), and
define two Fourier
transformed doublets of the Fermi field as
\begin{equation}
%5
\label{5}
a_{ik}=\int dx~ e^{-ikx}~q_{ix}~,~~~q_{ix}=\int d\widetilde k~
e^{ikx}~a_{ik}~,
\end{equation}
here $\widetilde k=k/(2\pi)$. When it is obvious from the context we omit
the spatial point designation (for example, it is $x$ in the
formula above). Then the
free Hamiltonian density and the density of interaction term may be presented
as
\begin{eqnarray}
%6
\label{6}
{\cal H}_0&=&\sum\limits_{k} k~a^+_{1k}a_{1k}-\sum\limits_{k}
k~a^+_{2k}a_{2k}~,\nonumber\\
[-.21cm]\\[-.2cm]
{\cal V}&=&2g~\sum\limits_{k}
a^+_{1k}a_{1k}\sum\limits_{l}a^+_{2l}a_{2l}~,\nonumber
\end{eqnarray}
(more precisely, the last expression should be symmetrized, but it does not
play a significant role for further). Fermion anticommutation relations lead,
as known
\cite{tir}, to the standard formulation of creation $a^+$ and annihilation
$a$ operators with the
standard anti-commutator valid $a^+ a~+~a a^+=1$ (indices are omitted).
It is easy to see that by the definition the reference state
corresponds to the eigenstate of Hamiltonian $H=\int dx {\cal H}$,
$H=H_0+V$ with zero eigenvalue $H|0)=0$, since $H_0|0)=0$, $V|0)=0$.
However, the states of free Hamiltonian with
negative energy (while working within the perturbation theory) give rise concern.
We see from Eq. (\ref{6}) that they are the states with negative momentum
$a_{1k}, k<0$ for particles of the first kind, and those are states with
positive momentum
$a_{2k}, k>0$ for particles of the second kind. Usually, this problem is
resolved by filling up
the Dirac sea with particles of negative energy. Thus, we take an ansatz
for ground state as follows
\begin{equation}
%7
\label{7}
|0\rangle=\prod\limits_{k\ge-\Lambda}^{-P} a^+_{1k}\prod\limits_{l\ge
P}^{\Lambda} a^+_{2l}~|0)~.
\end{equation}
We introduce some auxiliary cutoff momentum $\Lambda$ in Eq. (\ref{7}) that
should eventually be going to infinity. Besides, we introduce a boundary
momentum $P$, its
meaning becomes clear below. The system 'charge'\
$$Q=\int dx(q^+_1 q_1+q^+_2 q_2)=\sum\limits_{k}(a^+_{1k}  a_{1k}+
a^+_{2k} a_{2k})$$
commutes with Hamiltonian $H$ and, hence, it is convenient to classify
its eigenstates.

Then we have for the free Hamiltonian
$${\cal H}_0~ |0\rangle=\left(\sum\limits_{-\Lambda}^{-P}
k-\sum\limits^{\Lambda}_{P} k\right) |0\rangle~,$$
and for the interaction term
$${\cal V}~ |0\rangle=2gL (\Lambda-P)^2 |0\rangle~,$$
i.e. the energy density of such a Dirac sea looks like
\begin{equation}
%8
\label{8}
{\cal E}_{\mbox{\scriptsize{D}}}=-\Lambda (\Lambda+\widetilde 1)+P
(P+\widetilde 1)+2gL (\Lambda-P)^2~,
\end{equation}
where by definition the momentum unity is $\widetilde 1 = 2\pi/L$. It is
interesting to notice that the parabola branches as a function of the cutoff
$\Lambda$
for the coupling parameters $2gL>1$ change their directions, and the Dirac
sea may have even
a finite relative depth. Indeed, it makes sense to fill up the Dirac sea to
some point $P$ where
the Dirac sea is getting its minimal energy
\begin{equation}
%9
\label{9}
\Lambda-P_{\mbox{\scriptsize{min}}}=\frat12\frat{2\Lambda+\widetilde 1}{2gL+
\widetilde 1},~
{\cal E}_{\mbox{\scriptsize{min}}}=-\frat14\frat{2\Lambda+\widetilde
1}{2gL+\widetilde 1}
~(2\Lambda+\widetilde 1).
\end{equation}
Since we consider the system in a finite box it means an integer nearest to
this value of $P_{\mbox{\scriptsize{min}}}$. Besides, it is also interesting that the 'vacuum'\ state of Dirac
sea is degenerate
for almost all values of the coupling constant (there is an exact two-fold
degeneracy for
some discrete set of coupling constants) because the nearest integer either
exceeds or is less
than $P_{\mbox{\scriptsize{min}}}$. (It is clear that if this property is still valid for the
multi-dimensional
consideration, then the Dirac sea degeneracy is measured by the area of
corresponding sphere, see
below). We are talking about the relative depth of the Dirac sea in two-dimensional model
because pointing the parameters $\Lambda$ and $P$, being consistently related, at
infinity allows us to
reach the unlimited low values of ${\cal E}_{\mbox{\scriptsize{D}}}$.
Probably, it is rather the
specific feature of two-dimensional model (see an analysis of multi-dimensional model below).

We have a standard picture of the Dirac sea at low values of coupling
constant $gL\to 0$ but the boundary momentum $P$ becomes comparable with the
cutoff scale $P\sim
\Lambda/2$ already for the values of order $2gL \sim 1$. The excitations of
such a Dirac sea are
fairly curious. Adding or removing one particle of enormously huge momentum
$\sim \Lambda/2$
results in a small energy increase $\sim d {\cal E}/d
P|_{P=P_{\mbox{\scriptsize{min}}}}$ only, that,
apparently, is inconsistent with observations. Amazingly, such states assume
an existence of
particle separation mechanism as the particles of one kind acquire mainly
negative
momenta unlike the particles of another kind possessing the positive ones.
This behavior is rooted in a specific form of kinetic energy term of two-dimensional
model in Eq. (\ref{6})  if we consider the kinetic energy in the
nonrelativistic
approximation as a small deviation from the Fermi energy.

If we consider another example of ensemble with the same number of states
of positive and negative momenta for both types of particles then there
are the $L (\Lambda-P)$
particles of first type with positive momenta and the same number of
particles of this
type with negative momenta. Similar situation takes place for the particles
of second type. Then
the lowest energy for the particles of first type occurs if the states with
negative momenta
are collected from the sea bed (i.e. from cutoff momentum $\Lambda$). But the
states with
positive momenta fill the sea up starting from the lowest positive momenta,
i.e. from unit. Similar
picture takes place for the particles of second type with an obvious
permutation of states
with positive and negative momenta. Then we find the energy density as
\begin{eqnarray}
%10
\label{10}
&&{\cal E}_n=-\Lambda (\Lambda+\widetilde 1)+P (P+\widetilde 1)+
(\Lambda-P)(\Lambda-P+\widetilde 1)+\nonumber\\
&&+4gL (\Lambda-P)^2=-(\Lambda-P) 2P+4gL (\Lambda-P)^2~.
\end{eqnarray}
Comparing this result to Eq. (\ref{8}) we see 'neutral'\ ensemble energy
density is simply controlled by the total number of particles, but we can
show the 'absolute
depth'\ of its sea is poorly defined (tends to negative infinity). Then,
overall impression of
considering these particular examples suggests that the system properties
appear to be
dependent not only on the Hamiltonian but also on fixing the Hilbert space
sector (symmetries) adequate
to the problem considered \cite{fn}. It seems to us the similar results could
be hardly
received (in one-particle approximation) by simply calculating the
Hamiltonian
determinant only. Below we compare these results to those obtained for the
Thirring model
with point-like interaction $F(x)=\delta(x)$.

In order to complete this simple analysis we consider our system beyond the
chiral limit. Now the density of the free Hamiltonian takes the following
form
\begin{equation}
%11
\label{11}
{\cal H}_0=\sum\limits_{k} k~(a^+_{1k}a_{1k}-
a^+_{2k}a_{2k})+m(a^+_{1k}a_{2k}+a^+_{2k}a_{1k})~.
\end{equation}
Diagonalizing this form with a canonical transformation to new creation and
annihilation operators
\begin{eqnarray}
%12
\label{12}
&&\widetilde A_{1k}= \cos \varphi_m~a_{1k}+\sin \varphi_m~a_{2k}~,\nonumber\\
[-.21cm]\\[-.2cm]
&&\widetilde A_{2k}=-\sin \varphi_m~a_{1k}+\cos \varphi_m~a_{2k}~.\nonumber
\end{eqnarray}
where $\sin \theta_m=m/k$, $\cos \theta_m=k/k_0$, $k_0=(k^2+m^2)^{1/2}$,
$\theta_m=2\varphi_m$
we obtain the expansion of quark operators $q$ over  annihilation
operators (instead of Eq. (\ref{5})) as
\begin{equation}
%13
\label{13}
q_{jx}=\sum\limits_{k} e^{ikx}~[\widetilde U_k(j)~\widetilde A_{1k}
+\widetilde V_k(j)~\widetilde A_{2k}],
\end{equation}
here $j=1,2$. At $k>0$ the spinors have the form
\begin{eqnarray}
%14
\label{14}
&&\hspace{-0.5cm}\widetilde U_k(1)=
\left(\frat{k_0+k}{2k_0}\right)^{1/2}\!\!\!\theta_k,~
\widetilde U_k(2)= \left(\frat{k_0-k}{2k_0}\right)^{1/2}
\!\!\!\theta_k, \nonumber\\
[-.21cm]\\[-.2cm]
&&\widetilde V_k(1)=-\widetilde U_k(2)~~,~
\widetilde V_k(2)= \widetilde U_k(1)~,\nonumber
\end{eqnarray}
where $\theta_k$ is the theta-function ($\theta_k=1, k>0$, $\theta_k=0,
k\le0$).
Then, at $k<0$ we have similarly that
\begin{eqnarray}
%15
\label{15}
&&\hspace{-0.5cm}\widetilde U_k(1)=
~~\left(\frat{k_0+|k|}{2k_0}\right)^{1/2}\!\!\!\theta_{-k},~
\widetilde U_k(2)= -\left(\frat{k_0-k|}{2k_0}\right)^{1/2}
\!\!\!\theta_{-k},
\nonumber\\
[-.21cm]\\[-.2cm]
&&\widetilde V_k(1)=-\widetilde U_k(2)~,~~~
\widetilde V_k(2)=\widetilde U_k(1)~.\nonumber
\end{eqnarray}
It can be seen that the system 'charge'
$$Q=\int dx(q^+_1 q_1+q^+_2 q_2)=\sum\limits_{k}(\widetilde A^+_{1k}
\widetilde A_{1k}+
\widetilde A^+_{2k} \widetilde A_{2k})~,$$
commutes with the Hamiltonian $H$, and again it is convenient to classify
states in their 'charge'.
Dealing with the chiral limit $m=0$  we use notations $u_k(j)$, $v_k(j)$,
$j=1,2$, as the
ultimate expressions of formulae presented above. Now calculation of the
Dirac sea energy
density gives (instead of Eq. (\ref{8})) the following expression
\begin{equation}
%16
\label{16}
{\cal E}_{\mbox{\scriptsize{D}}}=-2\sum\limits_{k=P}^\Lambda k_0+2gL
(\Lambda-P)^2~,
\end{equation}
i.e., in principle, we obtain the picture quite similar to that we had in the
chiral limit up to the terms of order $O(m)$. The formulae are modified in a
similar way to
include small $O(m)$ corrections for the 'neutral'\ ensemble as well.

Now turning to another state which is analogous to the BCS state of paired
electrons in the superconductivity theory \cite{njl}, \cite{ff},
\cite{fujita} we analyse
again the chiral symmetric picture, first. Taking the zero value of ultimate
momentum $P$ for
the filled vacuum state in Eq. (\ref{7}) we perform the known canonical
transformation
\cite{ber}, introduce the particle operators $a$ for the states with positive
energy and
antiparticle operators $b$ for the states with negative energy
\begin{eqnarray}
%17
\label{17}
&&\hspace{-0.5cm}q_{1x}\!\!=\!\!\!\sum\limits_{k\ge
0}^\Lambda\!\!e^{ikx}a_{k}+
\sum\limits^{0}_{k\ge-\Lambda} \!\!\! e^{ikx} b^+_{-k}=\sum\limits_{k}\!
e^{ikx}(\widetilde\theta_k a_k+\widetilde \theta_{-k}b^+_{-k}),\nonumber\\
[-.21cm]\\[-.2cm]
&&\hspace{-0.5cm}q_{2x}\!\!=\!\!\!\sum\limits^{0}_{k\ge-
\Lambda}\!\!\!e^{ikx}a_{k}+
\sum\limits_{k\ge 0}^\Lambda \!\!\!e^{ikx} b^+_{-k}=
\sum\limits_{k}\!e^{ikx}(\widetilde \theta_{-k} a_k+\widetilde \theta_{k}
b^+_{-k}).\nonumber
\end{eqnarray}
Here $a^+$, $a$ и $b^+$, $b$ are the creation and annihilation operators of
quarks and antiquarks, $a|0\rangle=0$, $b|0\rangle=0$. It is also convenient
to use the
theta-function $\widetilde\theta_k$ at the appropriate intervals. Then the
density of free
Hamiltonian takes the form
\begin{equation}
%18
\label{18}
{\cal H}_0=\sum\limits_{k} k a^+_{k}a_{k}(\widetilde \theta_k-\widetilde
\theta_{-k})+
\sum\limits_{k} kb_{-k}b^+_{-k}(\widetilde \theta_{-k}-\widetilde
\theta_{k}).
\end{equation}
It is believed that the ground state of system at sufficiently intensive
interaction is formed by the quark--antiquark pairs with the opposite momenta
and vacuum quantum
numbers and is taken as a mixed state that is presented by the Bogolyubov
trial function (in that way
a particular reference frame is introduced)
$$|\sigma\rangle={\cal{T}}|0\rangle,~
{\cal{T}}=\prod\limits_{ p}\exp[\varphi_p~(a^+_{ p}b^+_{- p}+
a_{ p}b_{-p})].$$
The dressing operation ${\cal{T}}$ transforms the quark operators to the
creation and annihilation operators of quasiparticles
$A={\cal{T}}~a~{\cal{T}}^\dagger$,
$B^+={\cal{T}}~b^+{\cal{T}}^\dagger$. Now the representations (\ref{17})
becomes as
\begin{eqnarray}
%19
\label{19}
&&q_{1x}=\sum\limits_{k} e^{ikx}~(U_k A_k+U_{-k} B^+_{-k})~,\nonumber\\
[-.21cm]\\[-.2cm]
&&q_{2x}=\sum\limits_{k}\!e^{ikx}~(V_{k} A_k+V_{-k} B^+_{-k})~,\nonumber
\end{eqnarray}
in which the following designations are used
\begin{eqnarray}
%20
\label{20}
&&U_k~~=\cos\varphi~ u_{k}-\sin \varphi ~u_{-k}~,\nonumber\\
[-.21cm]\\[-.2cm]
&&U_{-k}=\sin\varphi ~u_{k}+\cos \varphi ~u_{-k}~.\nonumber
\end{eqnarray}
Similar formulae hold true for the components of $V$ with obvious
substitutions $U \to V$, $u \to v$. In order to unify the formulae
representation we introduce the
components $u$, $v$ (they are quite convenient at calculating beyond the
chiral limit) as
\begin{eqnarray}
%21
\label{21}
&&u_k=\widetilde \theta_{k}~,~~~~~~v_{k}=\widetilde \theta_{-k}~,\nonumber\\
[-.21cm]\\[-.2cm]
&&u_{-k}=\widetilde \theta_{-k}~,~~v_{-k}=\widetilde \theta_{k}~~.\nonumber
\end{eqnarray}
The free Hamiltonian density is transformed into
\begin{eqnarray}
%22
\label{22}
{\cal H}_0&=&\sum\limits_{k} k\cos\theta~ (A^+_{k}A_{k}-B_{-k}B^+_{-k})~
(\widetilde \theta_k-\widetilde \theta_{-k})+\nonumber\\
&+&\sum\limits_{k} k\sin\theta~ (A^+_{k}B^+_{-k}+B_{-k}A_{k})~
(\widetilde \theta_k-\widetilde \theta_{-k})~,
\end{eqnarray}
here $\theta=2\varphi$. The interaction term can be represented in the
following form
\begin{eqnarray}
%23
\label{23}
\hspace{-0.25cm}q^+_1 q_1 q'^+_2 q'_2&=&\sum\limits_{k}
(U^+_k A^+_k+U^+_{-k} B_{-k})(U_k A_k+U_{-k} B^+_{-k})\times,\nonumber\\
[-.21cm]\\[-.2cm]
&\times&\sum\limits_{l}(V^+_l A^+_l+V^+_{-l} B_{-l})(V_l A_k+V_{-l} B^+_{-
l}).\nonumber
\end{eqnarray}
As usual the pairing angle is calculated by minimizing the average energy
$\langle\sigma| H |\sigma\rangle$.
Due to the operator ordering accepted the nonzero contributions to this average
come only from the following matrix elements only
\begin{eqnarray}
&&\langle \sigma| B_{-k}B^+_{-k}|\sigma\rangle~,\nonumber\\
&&\langle \sigma| B_{-k}A_{k}~A^+_{l}B^+_{-l}~\widetilde F(k-
l)|\sigma\rangle~,\nonumber\\
&&\langle \sigma| B_{-k}B^+_{-k}B_{-l}B^+_{-l} ~\widetilde F(k-l)
\widetilde F(0)|\sigma\rangle~.\nonumber
\end{eqnarray}
We hold the form-factor in these notations in order to trace what are the
modifications necessary at considering a general form of interaction. The
first
contribution comes from the free Hamiltonian. The second matrix element
(remembering the form-factor
type $\widetilde F(k)=\delta(k)$ in the model we are interested in) leads to
the
following contribution
$$U^+_{-k} V_{-k} U_k V^+_k=-\frat{\sin^2\theta}{4}~(\widetilde
\theta_k+\widetilde \theta_{-k})~.$$
Both terms may lead to an energy gain unlike the contribution associated with
the third matrix element which is strictly positive.
However, as it was noticed in Ref. \cite{mz}, the third contribution vanishes
exactly
if the quark currents contain the generators of color gauge group $t^a$.
It results from calculating the trace over color group generators
of the tadpole diagrams,
and every contribution from the vertex exactly vanishes because of the spinor
basis completeness (here in a color space). Collecting all the contributions
together we obtain the
following expression for mean energy functional (trivial color factors are absorbed into coupling constant)
\begin{equation}
%24
\label{24}
\langle\sigma|{\cal H } |\sigma\rangle=\sum\limits_{k=0}^\Lambda
(-2 k \cos \theta -g~\sin^2\theta)~.
\end{equation}
The functional minimum is found by solving the following equation
\begin{equation}
%25
\label{25}
\sin \theta~(-k+g\cos\theta)=0~.
\end{equation}
Its non-trivial solution does exist for the momenta $k<g$ as it is seen from
$$\cos\theta=k/g~.$$
In order to keep the further steps transparent we are working  only with
those states and formally put the cutoff momentum as $\Lambda=g$. More
complicated
analysis with continuing this solution by using, for example, the trivial
branch $\theta=0$ can be done
but it is obviously superfluous. Calculating the condensate energy density
we have
\begin{equation}
%26
\label{26}
\langle\sigma| {\cal H }|\sigma\rangle=-\frat43~g^2~.
\end{equation}
This expression being compared to the energy density of 'neutral'\ system
(\ref{10}) shows that the 'neutral'\ ensemble energy at a certain magnitude
of ensemble density
becomes positive, i.e. the sea filling process by Bogolyubov states becomes
more profitable. As such
the condensate is characterized by the 'charge' density
\begin{equation}
%27
\label{27}
\langle\sigma| Q |\sigma\rangle=2gL~,~~Q=\int dx (q^+_1 q_1+q^+_2 q_2).
\end{equation}

The components of $U$, $V$ in
quark operators in the representation (\ref{10}) beyond the chiral limit
 has a form
\begin{eqnarray}
%28
\label{28}
&&\hspace{-0.35cm}U_k= \widetilde U_{k>0}(1)+\widetilde V_{k<0}(1)~,~
U_{-k}= \widetilde U_{k<0}(1)+\widetilde V_{k>0}(1)\nonumber\\
[-.21cm]\\[-.2cm]
&&\hspace{-0.35cm}V_k= \widetilde V_{k<0}(2)+\widetilde U_{k>0}(2)~,~
V_{-k}= \widetilde V_{k>0}(2)+\widetilde U_{k<0}(2)~.\nonumber
\end{eqnarray}
The canonical and dressing transformations are already performed with the
corresponding operators $\widetilde A_i$, i.e.
$$|\sigma\rangle={\cal{T}}|0\rangle,~
{\cal{T}}=\prod\limits_{ p}\exp[\varphi_p~(\widetilde A^+_{ p} \widetilde
B^+_{- p}+
\widetilde A_{ p}\widetilde B_{-p})],$$
$A={\cal{T}}~\widetilde A~{\cal{T}}^\dagger$, $B^+={\cal{T}}~\widetilde
B^+{\cal{T}}^\dagger$, (see Eq. (\ref{13})). The free Hamiltonian has a form
\begin{eqnarray}
%29
\label{29}
{\cal H}_0&=&\sum\limits_{k} k_0 \varepsilon(k)\cos\theta~ (A^+_{k}A_{k}-B_{-
k}B^+_{-k})~
(\widetilde \theta_k-\widetilde \theta_{-k})+\nonumber\\
&+&\sum\limits_{k} k_0 \varepsilon(k)\sin\theta(A^+_{k}B^+_{-k}+B_{-k}A_{k})
(\widetilde \theta_k-\widetilde \theta_{-k}).
\end{eqnarray}
Here, the function $\varepsilon(k)$ denotes a sign of momentum $k$. The
contribution of second matrix element is transformed into the following form
$$U^+_{-k} V_{-k} U_k V^+_k=-\frat{\sin^2(\theta-\theta_m)}{4}~
(\widetilde \theta_k+\widetilde \theta_{-k})~,$$
(the definition of auxiliary angle $\theta_m$ can be found in Eq.
(\ref{12})). As a result, the mean energy functional can be presented as
\begin{equation}
%30
\label{30}
\langle\sigma| {\cal H} |\sigma\rangle=\sum\limits_{k=0}^\Lambda
[-2 k \cos \theta -g~\sin^2(\theta-\theta_m)]~.
\end{equation}
In principle, it is not a great deal to show that we gain the minor
corrections of $O(m)$ only in two-dimensional consideration in comparison
to the results obtained in chiral limit. However, three-dimensional analysis
already shows the situation changes drastically \cite{mz}.

In order to compare the results obtained to the Thirring model (when
$F(x)=\delta(x)$)
in the chiral limit we should notice that the coupling constant is
dimensionless in
that model and differs from the coupling constant of the model with delta-like
form-factor in
the momentum space. Hereafter we are dealing with notations of Ref.
\cite{fujita}. As
known the respective Hamiltonian can be diagonalized by the Bethe
ansatz
\begin{eqnarray}
&&|k_1,\dots, k_N\rangle=\int \prod\limits_{i=1}^{N_1} d x_i e^{i k_i x_i}
\int \prod\limits_{j=1}^{N_2} d y_j e^{i k_{N_1+j} y_j}\times\nonumber\\
&&\times \prod\limits_{i, j}[1+\lambda_{ij} \epsilon(x_i-y_j)]
\prod\limits_{i=1}^{N_1} q^+_{1}(x_i)
\prod\limits_{j=1}^{N_2} q^+_{2}(y_j) |0\rangle~,\nonumber
\end{eqnarray}
where $\epsilon(x)$ is the step-like function $\epsilon(x)=-1$ at $x<1$ and
$\epsilon(x)=1$ at $x>1$, $k_i$ is the momentum of $i$-th particle,
$\varepsilon$ is the
infinitesimally small infrared regularizer and the phase factor looks like
$\lambda_{ij}=-g/2 S_{ij}$, $S_{ij}=(k_i E_j-k_j E_i)/(k_i k_j-E_i E_j-
\varepsilon^2)$,
$E_i$ is the particle energy (for the massless particles just under
consideration $E_i=|k_i|$). Now the equation for Hamiltonian eigenfunctions
is presented as
$$H |k_1,\dots, k_N\rangle =\sum\limits_{i=1}^{N} E_i~ |k_1,\dots,
k_N\rangle~,$$
where $N=N_1+N_2$. The periodic boundary conditions result in the
requirements for particle momenta
$$k_i=\frat{2\pi n_i}{L}+\frat{2}{L} \sum\limits_{j\ne i}^{N} \arctan (g
S_{ij}/2)~,$$
where $n_i=0, \pm 1,\dots, \pm N_0$, $N_0=(N-1)/2$. These conditions for the
'symmetric'\ vacuum state are obeyed for the following set of particle
momenta
\begin{eqnarray}
&&\hspace{-0.35cm} k_0=0~,~~(n_0=0), \nonumber\\
&&\hspace{-0.35cm} k_i=\frat{2\pi n_i}{L}+\frat{2 N_0}{L}\arctan (g/2),~~(n_i=1,2,\dots,
N_0),\nonumber\\
&&\hspace{-0.35cm} k_i=\frat{2\pi n_i}{L}-\frat{2 N_0}{L}\arctan (g/2),~~(n_i=-1,-2,\dots, -
N_0).\nonumber
\end{eqnarray}
Then the vacuum energy reads as
\begin{equation}
%31
\label{new_1}
E^{sym}_0=-\Lambda \left[N_0+1-\frat{2 N_0}{\pi} \arctan (g/2)\right]~,
\end{equation}
(the sign in front of the term containing an information on interaction can
be obtained by the continuity arguments basing on Eq. (\ref{8}), for
example). Taking
into account the definition of number of states as
$$N_0=\frat {L}{2\pi}~\Lambda~,$$
we can easily conclude that the result looks like an energy of ground states
Eq. (\ref{8}) if we remember the interrelation of energy density and ensemble
energy
${\cal E}=E/L$. It is worthwhile to notice that an interaction term can not
change the parabola signature
for the point-like interaction because of obvious limitation $|\arctan
x|<\pi/2$ as distinct
from the model with the delta-like form-factor in momentum space. (The similar
results take place
in the Neveu-Gross model \cite{al}.) It is known that for the massive Thirring
model the Dirac
sea distribution is
different from the free ($g=0$) one by renormalizing the rapidity
$\alpha \to \pi/(\pi g)\alpha$, $\alpha=\ln[(k_0-k)/m]$ only. The
requirement of finiteness of
physical excitation mass leads to the current mass renormalization
$m=c e^{- 2\varphi/(\pi+2\varphi)\Lambda}$, $\sin \varphi=g/(4+g^2)^{1/2}$,
and there appear the bound states in the spectrum of such a model. It was
shown in
Refs. \cite{ff}, \cite{fujita} that besides of 'symmetric'\ vacuum state
there exist
more profitable state in energy.

\section{Exact integrability of the KKB model}
Here, the behavior of quark ensemble with 'infinite' correlation length
is studied for the $3+1$ theory example that is
obviously of great
interest and all necessary modifications to be done at transiting to the
$D+1$-space are quite transparent. We start, first of all, with specifying the representation
of quark fields, they are
\begin{eqnarray}
%3_1
\label{3_1}
&&q_{{\vf x}}=\int d\widetilde {\vf p}
\frac{e^{-i{\vf p}{\vf x}}}{(2 p_0)^{1/2}}
\left[a_{{\vf p},s}u_{{\vf p},s}
+b^+_{-{\vf p},s}v_{-{\vf p},s}\right],\nonumber\\
[-.21cm]\\[-.2cm]
&&\bar q_{{\vf x}}=\int d\widetilde {\vf p}
\frac{e^{i{\vf p}{\vf x}}}{(2 p_0)^{1/2}}
\left[a^+_{{\vf p},s} \bar u_{{\vf p},s}
+b_{-{\vf p},s}\bar v_{-{\vf p},s}\right],\nonumber
\end{eqnarray}
here $\widetilde {\vf p}={\vf p}/(2\pi)^3$, the spinors $u$ and $v$ have a
standard form and normalizations conditions. Generally speaking, if one
follows two-dimensional
model of previous paragraph it is necessary to introduce the annihilation
(creation) operators
of the additional particle of different type instead of the creation
(annihilation) operators
$b$. However, here we introduce the particle and anti-particle operators
implying that the
corresponding canonical transformation with particles and holes has been
already done. This way is
convenient, as it becomes clear later, while dealing with the BCS state.
Besides, we need the
following commutation relation
\begin{equation}
%3_2
\label{3_2}
\{q_{i\alpha {\vf x}}, \bar q_{j\beta {\vf y}}\}=
\gamma^0_{\alpha\beta}~\delta_{ij}~\delta_{{\vf x},{\vf y}}~,
\end{equation}
and the interaction Hamiltonian in the following form
$$V=g~v~,~~v=\int d {\vf x} d {\vf y}~F({\vf x}-{\vf y})~j^a_\mu({\vf x})
j^a_\mu({\vf y})~,$$
where the current operators are meant as
\begin{equation}
%3_3
\label{3_3}
j^a_\mu({\vf x}) = \bar q_{\vf x}~\Gamma~q_{\vf x}~,
\end{equation}
with the compactifying (but a little bit inconsistent) designation
$\Gamma=\gamma_{\mu} t^a$.
It can be shown that Hamiltonian of the ensemble under consideration commutes
with its baryon charge
$$[H,Q]=0~,~~Q=\int d {\vf x}~\bar q_{{\vf x}}\gamma_{0}~q_{{\vf x}}~.$$
Thus, we can follow Thirring prescription, as at studying the two-dimensional
model, and assign a reference state $|0)$ that is annihilated by quark
operator
$$ q_{{\vf x}}|0)=0~,$$
at all respective discrete box points (i.e. all the states of antiparticles
described by $b$ type operators have been filled up) and the eigenstates of
Hamiltonian $H$
are sought in the following form
\begin{equation}
%3_4
\label{3_4}
|N) = \bar q_{{\vf z}_1} \bar q_{{\vf z}_2}\dots \bar q_{{\vf z}_N}~
\chi_{{\vf z}_1{\vf z}_2\dots {\vf z}_N}|0)~.
\end{equation}
The integration over all coordinates ${\vf z}_i$ is meant in this formula. It
can be shown that acting with the free Hamiltonian on such an eigenvector
results in a
superposition of the following form
\begin{equation}
%3_5
\label{3_5}
{\cal H}_0 |N)=\sum\limits_{{\vf k}} k_0 (a^+_{{\vf k}}
a_{{\vf k}}+b_{-{\vf k}}b^+_{-{\vf k}})|N)~,
\end{equation}
where $k_0=({\vf k}^2+m^2)^{1/2}$ is a quark energy. Similarly, it can be
received for the
interaction term
\begin{eqnarray}
%3_6
\label{3_6}
&&\hspace{-0.5cm}[v, \bar q_1 \bar q_2\dots \bar q_N] \chi=
2 N~ \bar q_1 \bar q_2\dots \bar q_N~\Gamma \gamma^0~ j~\chi+\nonumber\\
[-.21cm]\\[-.2cm]
&&\hspace{-0.5cm}+N(N-1)\bar q_1 \dots \bar q_{N-1}\Gamma \gamma^0 \bar
q_{N}\Gamma \gamma^0\chi
+N\bar q_1  \dots \bar q_N\Gamma \gamma^0 \Gamma \gamma^0\chi.\nonumber
\end{eqnarray}
This form needs to be elucidated because the corresponding permutations of
indices $1\dots N$ in the first and third terms, as well as all the
permutations of index pairs
in the second term are implied, but we omit all of them, just pointing out
that there appear $N$
or $N(N-1)$ equivalent contributions. Besides, we have also omitted the
spatial indices
of quark operators, and similarly the indices in quark current operator  $j$ of
Eq. (\ref{3_3}). As it
is constructed, the current operator acts on the reference state as $j|0)=0$.
This instrument
set allows us to find easily an action of the interaction operator on the
eigenvector
(\ref{3_4}). The direct product of $\lambda$-matrices that is present in the
second term
of Eq. (\ref{3_6}) can be decomposed into symmetric and anti-symmetric (over
color indices) parts
\begin{equation}
%3_7
\label{3_7}
{\vf \lambda} \otimes {\vf \lambda}=
\frat43~\Lambda_s-\frat83~\Lambda_a~,~~\Lambda_s+\Lambda_a=E_{\Lambda}~,
\end{equation}
where $E_{\Lambda}$ is a unit tensor. Similarly, the direct product of
spatial
${\vf \gamma}\gamma^0$-matrices,
\vspace{-0.2cm}
\begin{center}
\parbox[b]{2.75in}{
$%{3.6in}
{\vf \gamma}\gamma^0=
\left\| \begin{array}{rr}
{\vf \sigma} &0 \\
0&-{\vf \sigma}
\end{array}
\right\|,$
%\parbox[b]{2.5in}{$%{3.6in}
${\vf \gamma}\gamma^0=
\left\| \begin{array}{rr}
 0 & {\vf \sigma} \\
 {\vf \sigma} & 0
\end{array} \right\|,$
}
\end{center}
\vspace{-0.25cm}
\noindent
(in the chiral (Weyl), and standard representations, respectively) acts on
the spinor indices as a direct product of $\sigma$-matrices
\begin{equation}
%3_8
\label{3_8}
{\vf \sigma} \otimes {\vf \sigma}=\Sigma_s-
3\Sigma_a~,~~\Sigma_s+\Sigma_a=E_{\Sigma}~.
\end{equation}
Then,  we obtain for the direct product of $\gamma_\mu\gamma^0$-matrices
\begin{equation}
%3_9
\label{3_9}
\gamma_\mu\gamma^0 \otimes
\gamma^\mu \gamma^0=\Sigma_s+\Sigma_a-(\Sigma_s-3\Sigma_a)=4~\Sigma_a~.
\end{equation}
Here, the presence of symmetric and anti-symmetric projections in spinor
space is meant.
%
%%%%%%%%%%%%%%%%%%%%%%%%%%%%%%%%%%%%%%%%%%%%%%%%%
\begin{figure}%[!tbh]
\includegraphics[width=0.3\textwidth]{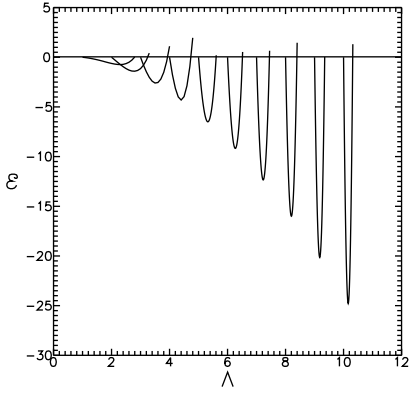}
\caption{The Dirac sea energy as a function of cutoff parameter $\Lambda$,
at $a =g~L^3= 1.025$ for several values of boundary momentum $P=1,\dots 10$.
}
\label{f1}
\end{figure}
%%%%%%%%%%%%%%%%%%%%%%%%%%%%%%%%%%%%%%%%%
As a result, we have for the antisymmetric over color and spinor indices
combination
\begin{equation}
%3_10
\label{3_10}
{\vf \lambda} \otimes {\vf \lambda}~~ {\vf \gamma}\gamma^0 \otimes
{\vf \gamma}\gamma^0=\frat{16}{3}~\Lambda_s~\Sigma_a~.
\end{equation}
Numerical factor in front of component $\Lambda_a \Sigma_s$ equals to zero.
The coordinate wave function is taken to be anti-symmetric. (Combining
$\Lambda_a$, $\Sigma_a$
with symmetric coordinate function corresponds to a repulsion being out of
our interest.)
The term where the product $\Gamma \gamma^0 \Gamma \gamma^0$ is available
gives for the
product of color matrices
\begin{equation}
%3_11
\label{3_11}
{\vf \lambda}~ {\vf \lambda}=2\frat{N_c^2-
1}{N_c}=\frat{16}{3}~E_{\Lambda}~(N_c=3).
\end{equation}
Similarly, we have for the product of ${\vf \gamma}\gamma^0$-matices
\begin{equation}
%3_12
\label{3_12}
\gamma_\mu\gamma^0 ~\gamma^\mu \gamma^0=-2~E_{\Sigma}~.
\end{equation}
Now we can calculate how the commutator Eq. (\ref{3_6}) acts on a reference
vector $|0)$ and receive
\begin{equation}
%3_13
\label{3_13}
[v, \bar q_1 \dots \bar q_N] \chi |0)=\!
\frat43 N \Bigl[ \Lambda_s \Sigma_a (N-3)-2\Lambda_a \Sigma_s\Bigr]|N).
\end{equation}
Considering this model in the chiral limit we evaluate the energy density
(negative one) coming from the free Hamiltonian contribution as
\begin{equation}
%3_14
\label{3_14}
{\cal E}_0=-2 N_c~ 4\pi \int\limits_{P}^\Lambda \frat{k^2 dk~k}{(2\pi)^3}=-
\frat{2N_c}{2\pi^2}~\frat14~(\Lambda^4-P^4)~,
\end{equation}
and the total number of particles with negative energy as
\begin{eqnarray}
%3_15
\label{3_15}
&&N={\cal N}~L^3~,\\
&&{\cal N}=
2 N_c~ 4\pi \int\limits_{P}^\Lambda \frat{k^2 dk}{(2\pi)^3}=
\frat{2N_c}{2\pi^2}~\frat13~(\Lambda^3-P^3)~,\nonumber
\end{eqnarray}
where ${\cal N}$ is the density of particles. These results allow us to write
down the Dirac sea energy in the form similar to Eq. (\ref{8}). For
simplicity, we suggest that the
combinations $\Lambda_s \Sigma_a$ and $\Lambda_a \Sigma_s$ contribute
identically, then
the energy density of the Dirac sea reads as
\begin{equation}
%3_16
\label{3_16}
{\cal E}_{\mbox{\scriptsize{D}}}={\cal E}_0+g~\frat43{\cal N}~(N-1)-
g~\frat{16}{3}{\cal N}~.
\end{equation}
%%%%%%%%%%%%%%%%%%%%%%%%%%
It is interesting (and we are sure, meaningful) that similar formula (as many other formulae
obtained above) has already merged in Ref.  \cite{atom}, see Eq. (3b) there).
%%%%%%%%%%%%%%%%%%%%%%%%%%
If the occupation numbers are large we may neglect the contribution of small
third term and, obviously, a unit in the second term. Now it is clear (it is
understandable from  the
dimensional analysis) that $a=g~L^3$ becomes an interaction parameter. Fig.
\ref{f1} shows the Dirac sea
energy as a function of parameter $\Lambda$ at $a=1.025$ for several values
of boundary momentum $P=1,\dots 10$.
Amazingly, it turns out that a relative depth of the Dirac
sea is finite as well, but unlike the two-dimensional model
it takes place at any value of $a$ parameter (the signature
of parabola changes at $a=2gL>1$ in the $1+1$ model).
An absolute depth of the Dirac sea is not defined (tends to
negative infinity at the cutoff parameter approaching positive infinity) just
in the same way as it
occurs in the two-dimensional model. However the presence
of term (related to the
interaction) with the highest (sixth) power of the cutoff parameter in Eq. (\ref{3_16}),
and the kinetic energy term
proportional to the fourth power leads to the energy distribution which is
getting a heavy narrowing with the
boundary momentum $P$ increasing and looks like a practically vertical line
in the limit (it is
seen in Fig. \ref{f1}). We might say that, actually, the Dirac sea is
reduced to a configuration
imaging a bound state.

%%%%%%%%%%%%%%%%%%%%%%%%%%%%%%%%%%%%%%%%%%%%%%%%%
\begin{figure}%[!tbh]
\includegraphics[width=0.3\textwidth]{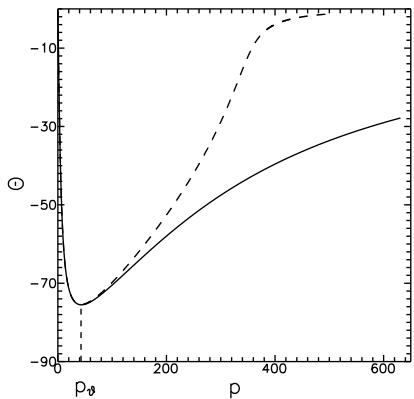}
\caption{The most stable equilibrium angles $\theta$ (in degrees) as a
function of momentum $p$ in MeV. The solid line corresponds to the NJL model
and the
dashed one corresponds to the KKB model.
}
\label{fangl}
\end{figure}
%%%%%%%%%%%%%%%%%%%%%%%%%%%%%%%%%%%%%%%%%
As the momentum, at which the minimum energy is reached, should be an integer
number (by construction), it is clear that a certain relation with the
coupling constant should spring up. Two real roots of the
equation ${\cal{E}}_{\mbox{\scriptsize{D}}}=0$ at
large
boundary momentum $P$ in the $D+1$-dimensional model are
\begin{eqnarray}
%3_17
\label{3_17}
&&\hspace{-0.25cm}\Lambda_1=P~,~~\Lambda_2=P+\Delta_\Lambda~,\\
&&\hspace{-0.25cm}\Delta_\Lambda \approx \frat{(D+1)~P}
{D[aD P^{D-1}-(D+1)/2]}\approx \frat{D+1}{a D^2}P^{2-D},\nonumber\\
&&a=\frat{D+1}{D^2}~2 N_c \frat{S_{\mbox{\scriptsize{D}}}}{(2\pi)^D}~gL^D~,~~
S_{\mbox{\scriptsize{D}}}=\frat{2\pi^{D/2}}{\Gamma(D/2)}~.\nonumber
\end{eqnarray}
 We propose to characterize the energy distribution by a width that
is defined by difference of this two roots $\Delta_\Lambda = \Lambda_2 - \Lambda_1$.
The minimal value of the
Dirac sea energy density is located approximately at
$\Lambda'\approx(\Lambda_1+\Lambda_2)/2$ and is given by
\begin{equation}
%3_18
\label{3_18}
\hspace{-0.35cm}{\cal{E}}_{\mbox{\scriptsize{D}}}(\Lambda')\approx
-\frat{(D+1)^2 P^{D+1}}
{4D\left[aD P^{D-1}\!\!-\!\!\frat{D+1}{2}\right]}
\approx-\frat{(D+1)^2}{4a D^2}P^{2}.
\end{equation}
The corresponding parabola (an enveloping of minimal energy points) and the
narrowing of energy distribution is clearly observed in Fig. \ref{f1}. The
distribution width is
constant for the $2+1$ dimensional model and in the $1+1$ dimensional model,
as we remember, the
width is proportional to the cutoff parameter $\Lambda$. The value of this
parameter at which the
distribution width becomes comparable with a minimal size of momentum
cell $2\pi/L$ can be considered
as a critical one $\Lambda_c$ because reaching this limit a degeneracy may
already become quite
essential. This parameter in the $3+1$-dimensional model is $\Lambda_c
\approx 2 N_c/(2gL)~2\pi/L$ (we have
singled out the dimensionless coupling constant $gL$ in this form because it
appears as a
natural theoretical parameter in next section). We have already faced such a
relation for colorless
interaction in the two-dimensional model.
Basing on the phenomenological estimate $g\sim$ 300 MeV
obtained in Ref. \cite{mz} and
taking into account the characteristic size of $L$ that is defined by
$\Lambda_{\mbox{\scriptsize{QCD}}}$
we may conclude that $\Lambda_c$ cannot be a large number.
%%%%%%%%%%%%%%%%%%%%%%%%%%
Actually, it looks like fairly justified to ask a question
whether a presence of critical momentum $\Lambda_c$
could signal a physical mechanism of cutting-off the
corresponding integrals.
%%%%%%%%%%%%%%%%%%%%%%%%%%

The point of real importance is that the states providing a relative minimum
of the Dirac sea are highly degenerate as an integer lattice of momenta (we
consider the quark
ensemble with periodic boundary conditions in a box of finite size $L$) does
not fit exactly the
sphere of radius $\Lambda'$. In two-dimensional model, as was discussed
above, the maximal degeneracy of
states forming the Dirac sea is two-fold only. It is easy to understand that
the degeneracy degree of
'vacuum'\ state at fixed cutoff parameter is proportional to the sphere area
of radius $\Lambda'$ at which
the minimum energy of the Dirac sea is reached, and it is going to infinity
at $P\to\infty$. It is well
known the energy of ensemble with a degenerate level, in principle, can be
reduced by removing a
degeneracy by introducing a breaking mechanism of state symmetry.
Another important point
to be taken into account is that energy distribution width tends to zero with
the cutoff value going to
infinity in the model of $3+1$ (or larger) dimensions. The vacuum energy
tends to a negative infinity
as $-\Lambda^2$,
%%%%%%%%%%%%%%%%%%%%%%%%%%
what, by the way, entails an interesting question how to define an antimatter state,
spectral representation, etc.
%%%%%%%%%%%%%%%%%%%%%%%%%%.
(One can see a full analogy with the results of conformal
theory \cite{cft}.)

These results show that the fluctuations (or removal of state degeneracy) will lead to
the destruction of such a
layer (infinitely thin). A state degeneracy could be also reduced by correlating
the pair states. We use
the well-known classification of such momentum distributions by dealing with
the total momentum of pair
${\vf P}$. The number of pairs with non-zero momentum, as known, is
proportional to a circumference
perimeter of two intersecting spheres with the characteristic radius
$\Lambda$ when their centers are located at the
distance of $|{\vf P}|$ from each other. The number of pairs at zero momentum
is much larger and is
proportional to the area of sphere with radius $\Lambda$. It is clear that
just such a subensemble
contributes dominantly. On the other hand, we know  that  there is an
attractive quark interaction in the
anti-triplet color channel and a diquark state can be more
beneficial. It suggests to consider
a diquark sea (a color superconductor state) as a vacuum ensemble and it
is not necessary to start
from the Dirac layer, as we see. This task and comparison with the BCS state
(strong interactions
phenomenology teaches this is a quite reasonable option) was examined in
detail in Ref. \cite{MZ3} and below
we use that information.  We recall, for the beginning, the BCS state is not
eigenstate of the
Hamiltonian and is a mixed state formed by the condensate of quark-antiquark
pairs, in the KKB model it was
studied in Ref. \cite{MZ2}.

The average specific energy per quark $w=E/(V\gamma)$ has been calculated in
the following form
\begin{equation}
%3_19
\label{w}
w=\int d \widetilde{\vf p}p_0(1-\cos\theta)
-\frat12\int d \widetilde{\vf p}
\sin \left(\theta-\theta_m\right)M({\vf p}),
\end{equation}
where $M({\vf p})$ is an induced quark mass
\begin{equation}
%3_20
\label{im}
M({\vf p})=2G\int d \widetilde{\vf q}~
\sin \left(\theta'-\theta'_m\right)~F_{{\vf p},{\vf q}}~,
\end{equation}
here the general form of form-factor is implied, $p_0=({\vf p}^2+m^2)^{1/2}$,
$\theta=2\varphi_{\vf p}$, $\varphi_{\vf p}$ denotes the pairing angle and
the primed variables (and
below as well) correspond to the integration over momentum ${\vf q}$, in
particular,
$\theta'=2\varphi_{\vf q}$. The $\theta_m$
angle is determined by relation $\sin \theta_m=m/p_0$. A unit in the first
term of Eq. (\ref{w}) is present because of normalizing to have the energy of
ground state equal zero
when an interaction is switched off. The most stable extremals of the
functional (\ref{w}) are
plotted in Fig. \ref{fangl} to compare the NJL model (solid line) to the KKB
model (dashed line) under
normal conditions ($T=\mu =0$).

%%%%%%%%%%%%%%%%%%%%%%%%%%%%%%%%%%%%%%%%%%%%%%%%%
\begin{figure}%[!tbh]
\includegraphics[width=0.3\textwidth]{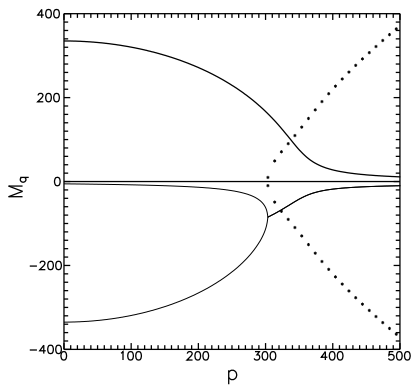}
\caption{Three branches of solutions for dynamical quark mass in MeV for
the KKB model as a function of momentum (MeV). The imaginary parts of
solutions are shown by dots.
}
\label{f2}
\end{figure}
%%%%%%%%%%%%%%%%%%%%%%%%%%%%%%%%%%%%%%%%%
The expression (\ref{w}) diverges for the delta-like form-factor in coordinate
space (the NJL model) and to obtain the reasonable results the upper limit
cutoff $\Lambda$ in the
momentum integration is introduced being one of the tuning model parameters
together with the
coupling constant $G$ and current quark mass $m$. Below we use one of the
standard sets of the parameters for
the NJL model $\Lambda=631$ MeV, $G\Lambda^2/(2\pi^2)\approx1.3$, $m=5.5$
MeV, whereas the
KKB model parameters are chosen in such a way that for the same quark current
masses the dynamical
quark ones in both NJL and KKB models coincide at vanishing quark momentum.
The momentum $p_\vartheta$
corresponds to the maximal attraction between quark and antiquark. The
inversed value of this parameter
determines a characteristic size of quasiparticle. It is of order of
$p_\vartheta \sim(m
M_q)^{1/2}$ (where $M_q$ is a characteristic quark dynamical mass) for the
models considered, i.e. the
quasiparticle size is comparable with the size of $\pi$-meson (Goldstone
particle). It is a
remarkable fact that the quasiparticle, as it is seen from Fig. \ref{fangl},
does not depend noticeably
on the form-factor profile or, in other words, on the scale, but rather
depends on the coupling
constant. Now we transform the expression for the specific energy (\ref{w})
into the form characteristic for
the standard mean-field approximation. Representing trigonometric factor in
the form of a certain
dynamical quark mass $M_q$
\begin{equation}
%3_21
\label{tv1}
\sin \left(\theta-\theta_m\right)=\frat{M_q}{P_0}~,~~
P_0=({\vf p}^2+M_q({\vf p}))^{1/2}~.
\end{equation}
and performing the algebraic transformations we can show that there is a
natural interrelation between the induced current and dynamical quark masses
\begin{equation}
%3_22
\label{mass}
M_q({\vf p})=M({\vf p})+m~,
\end{equation}
and the expression (\ref{w}) is transformed to
\begin{equation}
%3_22
\label{w2}
\hspace{-0.5cm}w=\int d \widetilde{\vf p}p_0-\int d \widetilde{\vf p}P_0
+\frat{1}{4G}\int d \widetilde{\vf p}d \widetilde{\vf q}F_{{\vf p},{\vf q}}
\widetilde M({\vf p})
\widetilde M({\vf q}),
\end{equation}
where $\widetilde M({\vf p})$ is the density of induced quark mass,
$P_0=[{\vf p}^2+M_q^2({\vf p})]^{1/2}$ is the energy of quark quasiparticle
with dynamical mass
\begin{equation}
%3_23
\label{3_23}
M_q({\vf p})=m+M({\vf p})=m+\int d \widetilde{\vf q}~
F_{{\vf p},{\vf q}}~\widetilde M({\vf q})~.
\end{equation}
In the particular case of the KKB model we have
\begin{equation}
%3_24
\label{mkeld}
M({\vf p})=2 G~\frat{M_q({\vf p})}{P_0}~.
\end{equation}
In practice, it is convenient to use an inverse function $p(M_q)$. In
particular, in the chiral limit $M_q=(4G^2-{\vf p}^2)^{1/2}$, at $|{\vf
p}|<2G$,
and $M_q=0$ at $|{\vf p}|>2G$. In this case the quark
states with momenta $|{\vf p}|<2G$ are degenerate in energy $P_0=2G$. Fig.
\ref{f2} demonstrates three branches of solutions of the equation (\ref{mkeld})
for dynamical quark mass.
The dots show the imaginary part of solutions which are generated at the
point where two real solution
branches are getting merged. The integrands in (\ref{w2}) are estimated as
follows:
$$p_0-P_0+\frat {1}{4G}~M^2 \sim -\frat{G~m^2}{p^2}~,$$
and, then we find a
linearly diverging integral for the specific energy of ensemble in the ($3+1$) dimensions
$$w\sim -\int \frat{dp~ p^2}{2\pi^2}~\frat{G~m^2}{p^2}~,$$
In the situation of $D+1$ dimensions the contributions are
proportional to $-\Lambda^{D-2}$. The integral converges for $D=1$ but there
is a logarithmic
divergence for $D= 2$. We have already
mentioned that in Eq. (\ref{w}) and Eq. (\ref{w2}) a simple regularization
was
used and to get specific energy density per quark the respective contribution,
that is proportional to $-\Lambda^{D+1}$ in the case of $D+1$ dimensions,
$$w_0=-\int d \widetilde{\vf p}~p_0~,$$
should be returned back.
Putting all together we see that we can definitely get a significantly
lower energy
$\sim -\Lambda^{D+1}$ ($D>1$) for the BCS state
than the contribution of the Dirac sea $\sim -\Lambda^2$. Correlation
contribution which comes from the terms containing $\sin(\theta-\theta_m)$ in
the functional
(\ref{w}) is significantly suppressed in comparison with the contribution
$\Lambda^{6}$,
($D=3$) of the Dirac sea ($\Lambda^{2D}$ for $D+1$
dimensions) and the problem of squeezing the energy distribution is
irrelevant for them. The BCS states are also preferable from the
phenomenological
point of view because they are characterized by a nonzero
chiral condensate which is finite in the chiral limit and diverging at $m\ne
0$ (nevertheless, the observable meson states are finite \cite{wemes}).
Similar analysis of diquark
states performed in \cite{MZ3} allows us to find a gap in the
anti-triplet
channel
(in the chiral limit)
$$\Delta=(4G_d^2-p^2)^{1/2}~,$$
with the energy $E=2G_d \approx$ 114 MeV. This energy is about three times
less than the quark energy in the BCS state (it is explained by a decrease of
coupling constant in
the anti-triplet channel). It has been demonstrated that the BCS state at
normal conditions of zero
temperature and zero baryon number density is more energy favorable
than the state of color superconductor.

Summarizing this section we may conclude that the degeneracy of ground state in the form
of Dirac sea could be a reason of vacuum state rearrangement because of
squeezing the energy distribution. We can already declare at this stage that  the ground state in
its traditional meaning does not, apparently, exist in such quark ensembles,
and  the corresponding systems are doomed, in a sense, to
fluctuate. Very similar problems in the theory of quantum phase transitions
and anomalous behavior of
Fermi-systems are discussed in the solid state physics.
The quantum phase transitions and anomalous behavior
of Fermi systems have been investigated very
actively in condensed matter physics and this process is still
going on \cite{amsti}.
%%%%%%%%%%%%%%%%%%%%%%%%%%
Here, we would like to note an amusing fact that the models of similar
Hamiltonian forms are widely used in physics of condensed matter,
nuclear physics while dealing with ensembles of finite particle number.
They are exactly integrable
\cite{r}, \cite{g}, \cite{c}
and well understood in the framework of conformal theory
 \cite{s}.
Seems, our results show the corresponding method could be a promising
one in our field as well.
%%%%%%%%%%%%%%%%%%%%%%%%%%

\section{Coupling constant}
\begin{figure}%[!tbh]
%%25
\includegraphics[width=0.3\textwidth]{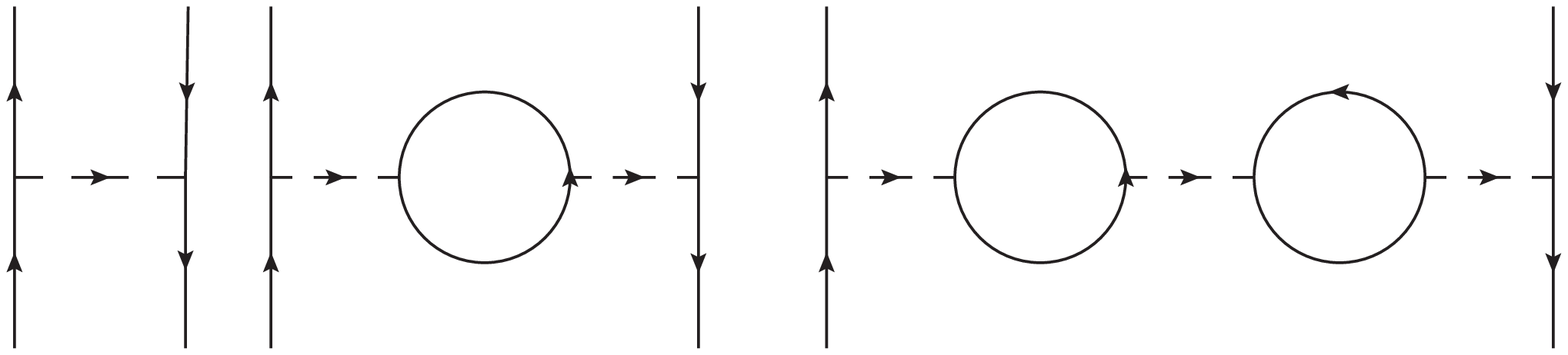}
\caption{The first terms of perturbative series for the observed
coupling constants.}
\label{frow}
\end{figure}

The KKB model provides us with yet another interesting opportunity to trace
back the interrelation between observed and bare coupling constants within an
entire energy interval. Despite its 'toy'\ form the KKB model is a field theory
with all the proper attributes including singularities. In particular,
it was noticed in Ref. \cite{wemes} that there exist the singular diagrams
(both ultraviolet and infrared divergent) in
addition to the regular diagrams that were used to calculate some results for the meson states
bound quarks.
Specifically, in the present paragraph we consider a number of diagrams that lead to the
modification of the bare coupling constant in scalar and pseudoscalar sectors,
%%%%%%%%%%%%%%%%%%%%%%%%%%%%%%%%%%%%%%%%%%%%%%%%%%%%%%%%%%%%%%%%%%%%%%%%%%%%%%%%
 (see Fig. \ref{frow}, where initial terms of a perturbative series are shown),
%%%%%%%%%%%%%%%%%%%%%%%%%%%%%%%%%%%%%%%%%%%%%%%%%%%%%%%%%%%%%%%%%%%%%%%%%%%%%%%%
because we need to control situation perturbatively as the BCS states
are not the eigenvalues of the Hamiltonian.
%%%%%%%%%%%%%%%%%%%%%%%%%%%%%%%%%%%%%%%%%%%%%%%%%
%%%%%%%%%%%%%%%%%%%%%%%%%%%%%%%%%%%%%%%%%%%%%%%%%
%%%%%%%%%%%%%%%%%%%%%%%%%%%%%%%%%%%%%%%%%%%%%%%%%
We present a four-fermion interaction as a product of two color
currents, taken at points ${\vf x}$ and ${\vf y}$  to underline its
nonlocal character in the Fig. \ref{frow}.
%%%%%%%%%%%%%%%%%%%%%%%%
%%%%%%%%%%%%%%%%%%%%%%%%
As was already mentioned above,
all the momentum integrations in the KKB model
are factorized, and an actual integration variable is only the (virtual)
quasiparticle energy. Then, the problem becomes in fact one-dimensional,
that is, seems to be a simplest one in this sense.
We further assume that quasiparticles with dynamical mass corresponding
to the momentum transfer ${\vf p}$, $M({\vf p})$ (in this
section we will for brevity use this notation for the quark dynamical mass)
take part in all virtual processes. This assumption (an approximation) seems to be
quite plausible if it is taken into account that in the KKB model the
quark dynamical mass, as shown, say, in Fig. \ref{f2}, smoothly transforms
into the mass of a bare (current) quark. It is not
hard to show that a perturbative series
%shown in Fig. \ref{frow}
can be
expressed in terms of the polarization operator
\begin{equation}
%55
\label{ap3_1}
\widetilde \Pi^{\pi,\sigma}=4 N_c \int \frat{d\widetilde {\vf k}}{E}
\frat{-2 E^2+M^2\mp M^2}{(\varepsilon^2/4)-E^2}
\end{equation}
(where $\varepsilon$ is the energy difference between the quarks in different spatial points,
${\vf x}$ and ${\vf y}$,
$E=[{\vf k}^2+M^2({\vf k})]^{1/2}$ stands for the energy of a loop quasiparticle,
and upper sign in the numerator corresponds to the pseudoscalar channel,
and the lower sign corresponds to the scalar one), as follows:
\begin{equation}
%56
\label{ap3_2}
G_o=G+VG\widetilde \Pi +(VG \widetilde \Pi)^2+\cdots,
\end{equation}
where $V$ is a volume the system is embodied in.
The system volume is exactly that infrared contribution we have mentioned
above. It appears to
be a consequence of nonlocality of the model and assumes the form of an
extra delta function $\delta^3(0)$. The standard regularization of the
contribution coming from the latter function leads just to the factor under
consideration. One can conclude from Eq. (\ref{ap3_1}) that the integral contains
a strong ultraviolet divergence. While discussing an expression for the
quark specific energy (\ref{w}), (\ref{w2}), we mentioned a natural way of
rendering
the formally divergent expressions sensible by normalizing them with
respect to the free Lagrangian (Hamiltonian). Let us do the same way (in
the spirit of renormalization theory) in the case at hand. To this end,
we assume that the observed polarization operator $\Pi$ is given by the
difference of $\widetilde\Pi$  and operator $\Pi_0$ generated by
the current quarks with mass $m$,
\begin{equation}
%57
\label{poobs}
\Pi=\widetilde\Pi-\Pi_0~.
\end{equation}
Since at large momenta the quark dynamical mass smoothly transforms into
the current one, it is clear that in the case of a fast enough
convergence to the quark current mass, any diagram of perturbation theory
will lead to a finite expression. In particular, in the chiral limit the
integrals are (automatically) strictly cutoff at the momentum $2G$ ($E=(4G^2-{\vf p}^2)^{1/2}$).
%Consider now in some detail how it happens for a set of diagrams shown in
%Fig. \ref{frow}.

Then, represent the expression (\ref{ap3_1}) as follows:
\begin{equation}
%58
\label{ap3_3}
\widetilde \Pi=N_c~(\widetilde I+\varepsilon^2 \widetilde J-
\widetilde K^{\pi,\sigma})~,
\end{equation}
where the following notations are used
\begin{eqnarray}
\label{ap3_4}
&&\widetilde I=2 \int d\widetilde {\vf k}~ \frat{1}{E}~,~~
\widetilde J=\frat12 \int d\widetilde {\vf k}~ \frat{1}{E}~
\frat{1}{E^2-\varepsilon^2/4}~,\nonumber\\
&&\widetilde K^\sigma=2 \int d\widetilde {\vf k}~ \frat{1}{E}
~\frat{M^2}{E^2-\varepsilon^2/4}~,~~K^\pi=0~.\nonumber
\end{eqnarray}
The mass in the KKB model is related with the energy by the relation $M-
m=2GM/E$. By taking into account the energy definition $E^2={\vf k}^2+M^2$,
the
momentum integral can be transformed in the energy one:
$$k dk= E dE~\left(1+\frat{2G~m^2}{(E-2G)^3}\right)~,$$
where
$$k=\frat{E}{E-2G}~\left[(E-2G)^2-m^2)\right]^{1/2}~.$$
The integrals $\widetilde I$, $\widetilde J$, $\widetilde K$
are calculabel in terms of elementary functions.
The first one is found to be
\begin{eqnarray}
\label{ap3_5}
\widetilde I&=&\frat{1}{\pi^2} \left[\left(\frat{\widetilde E_\Lambda}{2}
+2G-G \frat{m^2}{\widetilde E^2_\Lambda}\right) \widetilde s_\Lambda
+\right.\nonumber\\
&+&\left.\frat{4G^2}{3} \frat{\widetilde s^3_\Lambda}{\widetilde
E^3_\Lambda}
-\frat{m^2}{2}\ln \frat{\widetilde E_\Lambda+\widetilde s_\Lambda}{m}-
G m \arccos \frat{m}{E}~\right]~,\nonumber
\end{eqnarray}
where the following notation is used:
\begin{eqnarray}
&&\widetilde E=E-2G~,~~
E_\Lambda=\left[\Lambda^2+M^2(\Lambda)\right]^{1/2}~,\nonumber\\
&&\widetilde s_\Lambda=\left[(E_\Lambda-2G)^2-m^2\right]^{1/2}~,\nonumber
\end{eqnarray}
$\Lambda$ is a formal upper limit of momentum
integration. As we have already noted, physically meaningful results are
obtained if  $I_0$ calculated with a quark of bare mass
$$I_0=2 \int d\widetilde {\vf k}~ \frat{1}{e}=\frat{1}{\pi^2}
\left(\frat{e_\Lambda s_\Lambda}{2}-\frat{m^2}{2}\ln
(e_\Lambda+s_\lambda)\right)~,$$
 is subtracted, where $e=(k^2+m^2)^{1/2}$.
By taking the cutoff integration momentum to be  $\Lambda \gg G,m$
we expand the obtained expressions isolating a finite contribution
\begin{eqnarray}
\label{ap3_6}
\lim_{\Lambda \to\infty}I=\lim_{\Lambda \to\infty} \widetilde I- I_0
\to-\frat{1}{\pi^2}~\left(\frat{10}{3}~G^2
+\frat{\pi}{2}~Gm\right).\nonumber
\end{eqnarray}

For the integral $\widetilde J$ we have:
\begin{eqnarray}
\label{ap3_7}
&&\widetilde J=\frat{1}{4\pi^2}\frat{1}{\varepsilon}
\left[j_-i_--j_+i_++\sum\limits_{n=1}^4j_ni_n\right]~,\\
&&j_\pm=1-\frat{v_1}{2G_\pm}+\frat{v_2}{(2G_\pm)^2}
-\frat{v_3}{(2G_\pm)^3}+\frat{v_4}{(2G_\pm)^4},\nonumber\\
&&j_1=v_1 a_1-v_2 a_2+v_3 a_3-v_4 a_4~,\nonumber\\
&&j_2=v_2 a_1-v_3 a_2+v_4 a_3~,\nonumber\\
&&j_3=v_3 a_1-v_4 a_2~,\nonumber\\
&&j_4=v_4 a_1~,\nonumber\\
&&i_\pm=\int\limits_m^{E_\Lambda}
dE\frat{s}{E_\pm}~,~~i_n=\int\limits_m^{E_\Lambda}
dE\frat{s}{\widetilde E^n}~,\nonumber
\end{eqnarray}
where $E_\pm=E\pm\varepsilon/2$, $G_\pm=G\pm\varepsilon/4$,
$v_1=2G$, $v_2=0$, $v_3=2G m^2$, $v_4=4 G^2 m^2$,
$a_n=1/(2G_-)^n- 1/(2G_+)^n$,
(the terms containing $n=5$, $6$ will also appear in the
expression for $K^\sigma$). Divergent part of the integral $J^d$ is given by
the asymptotes of the following integrals:
\begin{eqnarray}
\label{ap3_8}
&&\hspace{-0.4cm}\lim_{\Lambda \to\infty}\int\limits_m^{E_\Lambda}
dE~\frat{s}{E_\pm}
\to E_\Lambda-
 2G_\pm \ln\frat{2E_\Lambda}{m}~,\nonumber\\
&&\hspace{-0.4cm}\lim_{\Lambda \to\infty}\int\limits_m^{E_\Lambda}
dE~\frat{s}{\widetilde E}
\to E_\Lambda~,~~
\lim_{\Lambda \to\infty}\int\limits_m^{E_\Lambda} dE~\frat{s}{\widetilde
E^2}\to
\ln\frat{2E_\Lambda}{m}~.\nonumber
\end{eqnarray}
The remaining terms in $\widetilde J$ are negligibly small compared with the
divergent ones. Using the definition of $a_n$ one can see that in the
asymptotic,
$\Lambda \to\infty$, the integral diverges only logarithmically:
$\lim_{\Lambda \to\infty} \widetilde J^d\to
\frat{1}{4\pi^2}\ln(2E_\Lambda/m)$.
We normalize results with respect to the free
Lagrangian. Being applied to the integral $\widetilde J$, this means that the
following contribution
$$J_0=\frat12 \int d\widetilde {\vf k}~ \frat{1}{e}~
\frat{1}{e^2-\varepsilon^2/4}~,$$
must be subtracted. For the divergent part, it is possible to obtain
$\lim_{\Lambda \to\infty} J_0^d\to
\frat{1}{4\pi^2}\ln\frat{2e_\Lambda}{m}$.
Considering that $E_\Lambda \to e_\Lambda$, when
$\Lambda \to\infty$ we see that
divergent parts in $J=\widetilde J- J_0$ are exactly mutually cancelled out.
For regular part one can derive:
\begin{eqnarray}
\label{ap3_10}
J^r&=&\widetilde J^r-J^r_0~,~~\widetilde J^r=\frat{1}{4\pi^2}
\frat{1}{\varepsilon}\biggl[j_-A_--j_+A_+-\nonumber\\
&-&j_1\left(2G+\frat{\pi m}{2}\right)-j_2+j_3~\frat{\pi}{4 m}+j_4 ~
\frat{1}{3m^2}\biggr]~,\nonumber
\end{eqnarray}
where the following notation is used:\\
\hspace{-0.5cm}$A_\pm=\!\!\left\{
\begin{array}{l}
-\left[(2G_\pm)^2-m^2\right]^{1/2} \ln\left|
\frat{2G_\pm-\left[(2G_\pm)^2-m^2\right]^{1/2}}{m}\right|,
\\
\!\!\left[m^2-(2G_\pm)^2\right]^{1/2}\!\!\left(\arcsin \frat{2G_\pm}{m}
-\arcsin \frat{2G_\pm+m}{|2G_\pm+m|}\right).
\end{array}
\right.
$
\\
the upper term is valid for $ |2G_\pm|\geq m$, the lower one, when
$|2G_\pm|< m$. For $J^r_0$ we have:
$$\hspace{-4.8cm}J^r_0=\frat{1}{4\pi^2}\frat{1}{\varepsilon}~\biggl(A^0_--
A^0_+\biggr)~,$$
\\
$A^0_\pm=\left\{
\begin{array}{l}
-\left[\frat{\varepsilon^2}{4}-m^2\right]^{1/2} \ln\left|
\frat{\pm\frat{\varepsilon}{2}-\left[\frat{\varepsilon^2}{4}-
m^2\right]^{1/2}}{m}\right|,
\\
\!\!\left[m^2-\frat{\varepsilon^2}{4}\right]^{1/2}\!\!\left(\arcsin
\frat{\pm\varepsilon}{2m}
-\arcsin
\frat{\pm\frat{\varepsilon}{2}+m}{\left|\pm\frat{\varepsilon}{2}+m\right|
}\right
)~.
\end{array}
\right.
$
\\
The upper term is valid when $|\varepsilon/2|\geq m$, the lower one
corresponds to the case
 $m>|\varepsilon/2|$. Similar results can be obtained for the integral
$K^\sigma=\widetilde K^\sigma-K^\sigma_0$.
The original integral $\widetilde K^\sigma$ can be represented in the form
analogous to Eq. (\ref{ap3_7})
\begin{eqnarray}
\label{ap3_11}
&&\widetilde K^\sigma=\frat{m^2}{\pi^2}\frat{1}{\varepsilon}
\left[k_-i_--k_+i_++\sum\limits_{n=1}^6k_ni_n\right]~,\nonumber\\
&&k_\pm=1-\frat{w_1}{2G_\pm}+\cdots+\frat{w_6}{(2G_\pm)^6}~,\nonumber\\
&&k_1=w_1 a_1-\cdots-w_6 a_6~,\nonumber\\
&&k_2=w_2 a_1-\cdots+w_6 a_5~,\nonumber\\
%&&k_3=w_3 a_1-\cdots-w_6 a_4~,\nonumber\\
%&&k_4=w_4 a_1-\cdots+w_6 a_3~,\nonumber\\
%&&k_5=w_5 a_1-w_6 a_2~,\nonumber\\
&&~~~~~~~~~~~~~~~~~ \cdots\nonumber\\
&&k_6=w_6 a_1~,\nonumber
\end{eqnarray}
where $w_1=6G$, $w_2=12 G^2$, $w_3=8G^3$, $w_4=12G^2 m^2$,
$w_5=24G^3m^2$, $w_6=16 G^4m^2$. Isolating regular part we have
\begin{eqnarray}
\label{ap3_12}
K_\sigma^r&=&\widetilde K_\sigma^r-K_{0\sigma}^r~,\nonumber\\
\widetilde
K_\sigma^r&=&\frat{m^2}{\pi^2}\frat{1}{\varepsilon}\biggl[k_-A_-
-k_+A_+-k_1\left(2G+\frat{\pi m}{2}\right)
-k_2+\nonumber\\
&+&k_3~\frat{\pi}{4 m}+k_4~\frat{1}{3m^2}+k_5~\frat{\pi}{16 m^3}+
k_6~\frat{2}{15 m^4}\biggr]~,\nonumber\\
K_{0\sigma}^r&=&\frat{m^2}{\pi^2}\frat{1}{\varepsilon}~
\biggl(A^0_--A^0_+\biggr)~.\nonumber
\end{eqnarray}
It is convenient to use dimensionless variables $m\to m/G$, $\varepsilon
\to \varepsilon/G$. One can see that the combination of the volume
and coupling constant of the form $VG^3$  is taken for the parameter in
theory, which characterizes the strength of the interaction. Generally speaking, it
is also obvious from the dimensional analysis.
%%%%%%%%%%%%%%%%%%%%%%%%%%%%%%%%%%%%%%%%%%%%%%%%%
\begin{figure}%[!tbh]
%26
\includegraphics[width=0.3\textwidth]{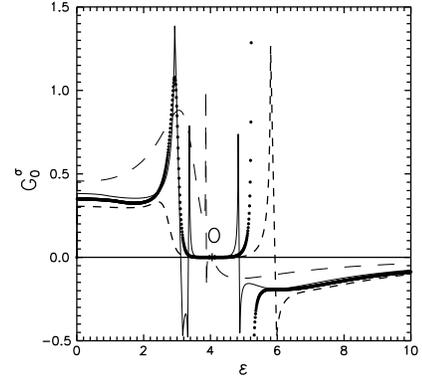}
\caption{The observed coupling constant $G_o^\sigma$ as a function of
the transferred energy $\varepsilon$ (in dimensionless variables). The short dash
line is obtained for $m = 0.9$. Dots show the case $m = 0.6$. The solid line
corresponds to the case $m = 0.4$. Everywhere $V = 1$. Point $O$ shows the
location where $G_o^{\pi,\sigma}=0$. The long dash line
corresponds to $m = 0.01$.}
\label{fsiren}
\end{figure}
%%%%%%%%%%%%%%%%%%%%%%%%%%%%%%%%%%%%%%%%%%%%%%%%%
The observed coupling constant $G_o$ for each individual channel separately
is expressed via a regularized polarization operator:
\begin{eqnarray}
%59
\label{ap3_13}
G_o^{\pi,\sigma}=\frat{G}{1-VG^3 \Pi^{\pi,\sigma}}~.
\end{eqnarray}
The polarization operator $\Pi$ in this expression is presented in the
dimensionless form; initially, it is proportional to the coupling
constant squared: $\Pi\sim G^2$. So, the polarization operators introduced
are free of typical logarithmical singularities and do not feature any
divergent parts at all. For definiteness, consider the positive
transferred energy $\varepsilon>0$ (the case of negative transferred energy
is clearly symmetric). From the formulae presented it follows that the
polarization operator contains strong pole singularities at the energy
value $\varepsilon=4$ (in dimensional units $\varepsilon=4G$),
where the variable $G_-$ vanishes.
Local vicinity of this point, as well as actually all obtained
expressions, deserves to be thoroughly studied analytically. However, in
order to simplify a discussion, we will limit
ourselves to carrying out a brief qualitative analysis and present for
illustrative purposes a number of figures. The pole singularities (of
maximal power $4$ for the integral  $\widetilde J$ and $6$ in
$\widetilde K$) lead to the observed coupling constant $G_o^{\pi,\sigma}$
vanishing at the energy value $\varepsilon=4$. It is clear that perturbation theory is valid
 in the vicinity of this point.

\begin{figure}%[!tbh]
%27
\includegraphics[width=0.3\textwidth]{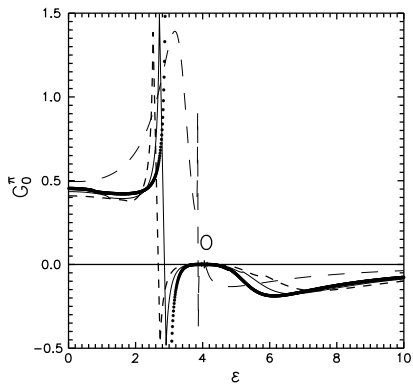}
\caption{The observed coupling constant $G_o^\pi$ as a function
of the energy $\varepsilon$. Notations
are the same as for the scalar channel of Fig. \ref{fsiren}.}
\label{fpiren}
\end{figure}

It can be shown that the bound states, defined by
the denominator poles of Eq.  (\ref{ap3_13}) (as all the remaining contributions are
negligibly small being compared with the pole singularities), may show up
in this region, and a pole in the denominator may appear either  right or  left
of the region under discussion (in dependence on the
sign of high order pole singularities).
Figures \ref{fsiren}, \ref{fpiren}
shows the energy $\varepsilon$ dependence of observed coupling constants $G_o^\sigma$,
$G_o^\pi$ respectively, (in dimensionless variables).
The dashed curves are obtained at $m = 0.9$, while the dot curves
present the case
$m = 0.6$, and the solid lines correspond to  $m = 0.4$. The system volume of the system
is assumed to be a $V = 1$, for definiteness. The location
of point in which the observed coupling constant vanishes:
$G_o^{\pi,\sigma}=0$ is marked as $O$. Now comparing the curves, one can clearly see the evolution
of some new ('resonance'\ ) state, that is manifested by a sharp and
sufficiently broad peak and is transformed into a bound state, with the parameter $m$
changing from the value $m = 0.9$ to $m = 0.4$. (The pole
singularities in both figures are somewhat smoothed in drawing in order not to lose
the regular 'resonance'\ structures.) In the  sigma-channel
several bound states simultaneously appear, when the parameter $m$ is
decreasing (see a respective curve with $m = 0.4$). Both figures  show also
the version with the parameter $m$ decreasing to the values specific
for the NJL model, that is, of order $m \sim 0.01$, the corresponding data are shown
by the long-dash curve. Both figures expose definitely not all the bound states.
Some of those are so narrow that it is impracticable to depict them
projecting on the scale used in the figures. This result makes, in principle,
possible to observe a transformation of the resonance into a genuine bound state.
The $m$ dependence becomes more pronounced  with $VG^3$
parameter increasing. The observed coupling
constant is substantially reduced in a low-energy region,
demonstrating a transition to the qualitatively different scale.
Overall, it can be concluded that if the parameter $VG^3$
regulating the interaction strength is small ($VG^3 < 10$), then the
energy dependence of observed coupling constant
is sufficiently smooth up to  values $\varepsilon\sim 3$--$4$,
and beyond it the bound states are coming into the play.
%%%%%%%%%%%%%%%%%%%%%%%%%%%%%%%%%%%%%%%%%%%%%%%%%
%%%%%%%%%%%%%%%%%%%%%%%%%%%%%%%%%%%%%%%%%%%%%%%%%
As a consequence, an adequate picture of
spontaneous breaking of chiral symmetry can be developed at this scale
providing us with a reasonable information of meson observables
and plausible scenario of diquark condensation.

\section{Conclusions}
%1
Throughout our work we have seen that picture of the ground state
of fermions (quarks) ensemble developing strong correlations
may significantly differ sometimes from
standard scenario (accepted intuitively) of exchange interaction, obtained from our
everyday experience (from condensed matter physics).
The Dirac sea displays a finite relative depth, that decreases
as $-\Lambda^2$, with momentum cutoff $\Lambda$ increasing.
There exist some critical value of the coupling constant $g_1$ in two dimensional model
 and when it is weak $g<g_1$ the Dirac sea becomes a standard one.
The Dirac sea width at $g>g_1$ increases linearly with momentum cutoff $\Lambda$ increasing.
In the three dimension situation the width is constant.
The Dirac sea width, as we show, becomes squized behaving as
$\Delta_\Lambda\sim\frat{D+1}{gL^D D^2}\Lambda^{2-D}$ for spatial ($D$) dimensions.
Coincidence Dirac sea width with a size of elementary cell
$\widetilde 1=2\pi/L$ (D=3)
specifies a critical value of momentum cutoff as $\Lambda_c \approx 2 N_c/(2gL)~2\pi/L$.
The existing estimates teach the parameter $\Lambda_c$ should not develop
a macroscopical value.
We demonstrate the Dirac sea is strongly degenerate with respective degeneracy power
proportional to the surface of ($D-1$)-dimensional sphere
Clearly, such a ground state is highly unstable and
according to a philosophy the Jahn--Teller theorem its energy could be lowered
by reducing of its symmetry (hence removing degeneracy).
It is difficult to free ourselves  from an idea that similar ground state
resemblance strongly the 'Big Bang'\ scenario.
%2
Plausible mechanism of degeneracy removal (vacuum state reconstruction)
with nonabelian (color) interaction switched on
could be seen as the Bogolyubov like state
of coupled quark antiquark pairs with zero total momentum
and vacuum quantum numbers. The energy of such a state is getting
a minimal magnitude in average.
(A technical reason of this feature appearance is rooted in
vanishing the contribution of tadpole diagrams.)
%3
It is interesting to notice that the models considered
despite the seemingly toy form possess all the attributes of quantum field
theory, including divergence.
It can be seen that there are strongly singular diagrams
 in the intermediate perturbation theory calculations,
but in final expressions there is not
any trace of the divergences and, seems, the general lanscape of this theory is determined
by the scenario of the Dirac sea filling.

Eventually, we would like to notice that several 
intuitive arguments of recent interesting development in favor of 
a confined quarkyonic phase existing \cite{larry}
receive surprisingly almost exact theoretical substantination in theframework of
our consideration (of course, if deconfinement paradigm is replaced).

We are thankful to  Prof. L. V. Keldysh for his insightful 
comments and deeply indebted to B. A. Arbuzov, S. B. Gerasimov, L. McLerran,
A. M. Snigirev, L. Turko, I. P.  Volobuev for fruitful discussions.
The work was supported by the State Fund for Fundamental
Research of Ukraine, Grant \nomer{Ph58/04}.

%\newpage


\begin{thebibliography}{99}
%1
\bibitem{fad}
L. D. Faddeev, Theor. Mat. Fiz. {\bf 148} (2006) 133.
%2
\bibitem{tir}
W. Thirring, Ann. Phys. {\bf 3} (1958) 91;\\
J. M. Luttinger, J. Math. Phys. {\bf 4} (1963) 1154;\\
D. C. Mattis and E. H. Lieb, J. Math. Phys. {\bf 6} (1965) 304.
%3
\bibitem{bik}
"Integrable Quantum Field Theories"\ , Lecture Notes in Physics, Vol. 151,
Springer-Verlag, New York, 1982, ed. by J. Hietarinta and C. Montonen;\\
N. M. Bogolyubov, A. G. Izergin, V. E. Korepin,
"Correlation function of integrable systems
and quantum inverse scattering method"\ , Nauka, Moscow, 1992;\\
V. Mastropietro and D. C. Mattis, "Luttinger model.
The First 50 Years and Some New Directions"\ ,
Series on Directions in Condensed Matter Physics---Volume 20, 2014;\\
A. B. Zamolodchikov, Al. B. Zamolodchikov,
"Conformal Field Theory and Critical Phenomena in Two-Dimensional Systems"\ .
%4
\bibitem{kkb}
M. V. Sadovskii, "Diagrammatics"\ , Singapore: World Scientific, 2006.\\
L. V. Keldysh, Doctor thesis, FIAN, (1965);\\
E. V. Kane, Phys. Rev. {\bf 131} (1963) 79;\\
V. L. Bonch-Bruevich, in "Physics of solid states"\ , M., VINITI, (1965).
%5
\bibitem{mz}
G. M. Zinovjev  and S. V. Molodtsov, Theor. Mat. Fiz.  {\bf 160} (2009) 444;\\
S. V. Molodtsov and G. M. Zinovjev,  Phys. Rev. {\bf D 80} (2009) 076001;\\
S. V. Molodtsov, A. N. Sissakian and G. M. Zinovjev,
Europhys. Lett. {\bf 87} (2009) 61001.
%6
\bibitem{atom}
B. F. Bayman, Nucl. Phys. {\bf 15} (1960) 33.
%7
\bibitem{ato}
J. M. Blatt, Prog. Theor. Phys. {\bf 24} (1960) 851;\\
K. Dietrich, H. J. Mang and H. Pradal, Phys. Rev. {\bf 135} (1964) B22.
%8
\bibitem{sim}
Yu. A. Simonov,  Phys. Lett. {\bf B 412} (1997) 371.
%9
\bibitem{fn}
John von Neumann, "Mathematical Foundation of Quantum Mechanics"\ ,
Princeton University Press, 1955.
%10
\bibitem{njl}
Y. Nambu and G. Jona-Lasinio, Phys. Rev. {\bf 122} (1961) 345.
%11
\bibitem{ff}
M. Faber and A. N. Ivanov, Eur. Phys. J. {\bf C 20} (2001) 723;
Phys. Lett. {\bf B563} (2003) 231;\\
T. Fujita, M. Hiramoto, T. Homma, and H. Takahashi,
J. Phys. Soc. Jap. 74 (2005) 1143.
%12
\bibitem{fujita}
T. Fujita, M. Hiramoto and H. Takahashi, "Bosons after symmetry breaking in
quantum field theory"\ , Nova Science Publishers, Inc. New York, 2009.
%13
\bibitem{ber}
F. A. Berezin, "The second quantization method"\ , Nauka, Moscow, 1986;\\
J.-P. Blaizot and G. Ripka, "Quantum theory of Finite Systems"\ , The MIT Press, 1985.
%14
\bibitem{al}
N. Andrei and J. H. Lowenstein, Phys. Rev. Lett. {\bf 43} (1979) 1698.
%15
\bibitem{cft}
F. C. Alcazar, M. N. Barber, and M. T. Batchelor, Ann. Phys. {\bf 182} (1988) 280;\\
J. L. Cardy, J. Phys. {\bf A 17} (1984) L385;\\
H.W.J. Bl\"ote, J. H. Cardy, and M. P. Nightingale, Phys. Rev. Lett. {\bf 56} (1986) 742;\\
I. Affleck, Phys. Rev. Lett. {\bf 56} (1986) 746.
%16
\bibitem{MZ3}
S. V. Molodtsov and G. M. Zinovjev, arXiv:1311.6606 [hep-ph].
%17
\bibitem{MZ2}
S. V. Molodtsov and G. M. Zinovjev, Europhys. Lett. {\bf 93} (2011) 11001;\\
S. V. Molodtsov and G. M. Zinovjev, Phys. Rev. {\bf D 84} (2011) 036011;\\
G. M. Zinovjev  and S. V. Molodtsov, Yad. Fiz. {\bf 75} (2012) 262.
%18
\bibitem{wemes}
G. M. Zinovjev and S. V. Molodtsov,  Yad. Fiz. {\bf 75} (2012) 1387;\\
G. M. Zinovjev, M. K. Volkov and S. V. Molodtsov,
Theor. Mat. Fiz.  {\bf 161} (2010) 408.
%19
\bibitem{amsti}
V. R. Shaginyan, M. Ya. Amusia and K. G. Popov,
Usp. Fiz. Nauk. {\bf 177} (2007) 585;\\
S. M. Stishov, Usp. Fiz. Nauk. {\bf 174} (2004) 853;\\
S. Sachdev, "Quantum phase transitions"\ , Cambridge University Press, 1998.
%20
\bibitem{r}
R. W. Richardson, Phys. Lett. {\bf 3} (1963) 277;\\
R. W. Richardson and N. Sherman, Nucl. Phys. {\bf B52}, (1964) 221;\\
R. W. Richardson, J. Math. Phys. {\bf 6} (1965) 1034.
%21
\bibitem{g}
M. Gaudin, J. Physique {\bf 37} (1976) 1087.
%22
\bibitem{c}
M. C. Cambiaggio, A. M. F. Rivas and M. Saraceno, Nucl. Phys. {\bf A 624} (1997) 157.
%23
\bibitem{s}
G. Sierra, Nucl. Phys. {\bf B572} (2000) 517;\\
J. Dukelsky, S. Pittel and G. Sierra, Rev. Mod. Phys. {\bf 76} (2004) 643.
%24
\bibitem{larry}
L. Mclerran, R. D. Pisarski, Nucl. Phys. {\bf A796} (2007) 83;\\
L. Mclerran, arXiv:1105.4103 [hep-ph].
\end{thebibliography}
\end{document}